\documentstyle[epsf,twocolumn,aps]{revtex}

\newcommand{\be}{\begin{equation}}
\newcommand{\ee}{\end{equation}}
\newcommand{\ba}{\begin{array}}
\newcommand{\ea}{\end{array}}

\newcommand{\bac}{\begin{array}{c}}
\newcommand{\bal}{\begin{array}{l}}
\newcommand{\baR}{\begin{array}{r}}
\newcommand{\bacc}{\begin{array}{cc}}
\newcommand{\ball}{\begin{array}{ll}}
\newcommand{\balr}{\begin{array}{lr}}
\newcommand{\barl}{\begin{array}{rl}}
\newcommand{\baccc}{\begin{array}{ccc}}
\newcommand{\barcl}{\begin{array}{rcl}}
\newcommand{\balcl}{\begin{array}{lcl}}
\newcommand{\barcll}{\begin{array}{rcll}}
\newcommand{\barll}{\begin{array}{rll}}
\newcommand{\barrclcl}{\begin{array}{rrclcl}}
\newcommand{\bacl}{\begin{array}{cl}}
\newcommand{\bacll}{\begin{array}{cll}}
\newcommand{\eac}{\end{array}}
\newcommand{\ber}{\begin{eqnarray}}
\newcommand{\eer}{\end{eqnarray}}

\newcommand{\half}{\frac{1}{2}}

\newcommand{\p}{^{\prime}}
\newcommand{\pp}{^{\prime\prime}}

\renewcommand{\Im}{{{\mathcal{I}}m}}

\begin{document}
\twocolumn[\hsize\textwidth\columnwidth\hsize\csname @twocolumnfalse\endcsname

\title{The Josephson Effect in Single Spin Superconductors}

\author{Robert E. Rudd$^{\dag}$ and Warren E. Pickett$^{\ddag}$}

\address{Naval Research Laboratory, Washington, DC 20375-5345}
\address{$^\dag$ SFA, Inc, 1401 McCormick Drive, Largo MD 20774}
\address{$^{\ddag}$ Department of Physics, University of California,
Davis CA 95616}

\date{September 12, 1997}
\maketitle

\begin{abstract}
The Josephson Effect provides a primary signature of single spin
superconductivity (SSS), 
the as yet unobserved superconducting state which was proposed recently
as a low temperature phase of half-metallic antiferromagnets.
These materials are insulating in the spin-down channel
but are metallic in the spin-up channel.
The SSS state is characterized by a unique form of 
p-wave pairing within a single spin channel.
We develop the theory of a rich variety of Josephson effects that
arise due to the form of the SSS order parameter.
Tunneling is allowed at a SSS-SSS' junction but of course
depends on the relative orientation of
their order parameters.  No current flows
between an SSS and an
s-wave BCS system due to their orthogonal symmetries, which
potentially can be used to 
distinguish SSS from other superconducting states.  Single spin
superconductors also offer a means to probe other materials, where
tunneling is a litmus test for any form of ``triplet'' 
order parameter.  
\vspace{5mm}
\end{abstract}

]


Recently one of the authors has used local spin density
functional calculations to identify a few good candidates
for half-metallic antiferromagnets (HM AFMs).\cite{HMAFM} 
Half-metallic (HM) materials have the property that charge transport 
is 100\% spin polarized: one spin channel is metallic (chosen
as `up' by convention),
while the down channel is insulating.  HM {\it anti}ferromagnets
are distinguished from the HM ferromagnets by having no 
macroscopic magnetization, yet charge transport is still 100\%
polarized.  A pairing instability in a HM AFM leads to 
superconductivity in only the metallic channel, a condensation
that has been called single spin superconductivity (SSS).\cite{SSS1,SSS2}
We note briefly the special properties of these materials, and then
discuss effects that can be observed in tunneling.

Several HM {\it ferro}magnetic materials \cite{DG1,IK} are strongly
indicated from theoretical studies, such as 
CrO$_2$ \cite{KHS},
the Heusler alloys UNiSn and NiMnSb \cite{DG1} and probably the colossal
magnetoresistance manganates \cite{WEPDJS}.
The properties of some of these materials have 
supported the HM behavior\cite{Fujii,hanssen}. 
Since HM materials have the remarkable property that (neglecting
spin-orbit coupling) the spin
moment is quantized, a necessary condition for HM antiferromagnetism
is that magnetic ions within the unit cell have antialigned moments
which cancel.  This criterion can be applied to limit the thousands
of possible magnetic compounds within even a given crystal structure,
to a small class that can be studied individually.
Ref.\ \cite{HMAFM} identifies, within the double perovskite
structural class La$_2M\p M\pp $O$_6$, the
two candidates La$_2$VCuO$_6$ and La$_2$MnVO$_6$, 
from a search of six pairs of transition metal ions giving a double
perovskite structure.  The MnV material is more likely to be an
exotic superconductor (See Ref.\ \cite{HMAFM}).
A third compound, La$_2$MnCoO$_6$, also
has a HM AFM phase but it is clearly unstable to ferromagnetism.
The density of states for the HM AFM phase of La$_2$VCuO$_6$
is shown in Fig.\ 1.  

\begin{figure}[tbp]
\epsfysize=6cm\centerline{\epsffile{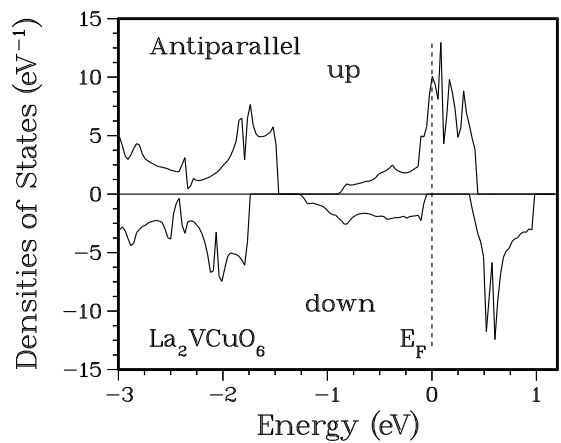}}
\caption{Total densities of states for each spin direction in the
double perovskite compound La$_2$VCuO$_6$.[1]  E$_F$ denotes
the Fermi level.  Note that, while
equal numbers of states are occupied in each spin channel,  
only the up channel is metallic.
\label{fig-calcDOS}}
\end{figure}

The proposed SSS phase\cite{SSS1,SSS2}
is a new kind of superconductor,
with unique properties arising due to the feature that only spin up
electrons participate in pairing.  A Cooper pair 
has total spin $S=1$, something familiar from superfluid
$^3$He and some theories of p-wave pairing in heavy fermion
superconductors.  Indeed, the systems have many common features, but
the fact that only one spin component is present in SSS leads to
important distinctions, specifically, $S_z=1$ and the supercurrent
is 100\% polarized.  In effect, the spin quantum variable
is irrelevant to the
bulk superconducting ground state.  Fermi statistics require that the 
orbital pair wave function is odd under exchange, characteristic
of the triplet form of pairing.


In Ref.\ \cite{SSS2} we compared and contrasted the SSS phase with
the conventional BCS theory of pairing, and enumerated
the symmetries of all allowed
order parameters for cubic, tetragonal, and hexagonal crystal
lattices.  The phase diagram for the cubic lattice is shown
in Fig. 2.

\begin{figure}[tbp]
\epsfysize=7cm\centerline{\epsffile{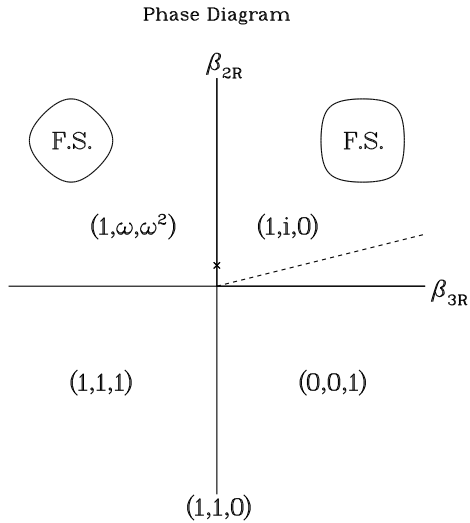}}
\caption{The Ginzburg-Landau phase diagram for SSS with $\Gamma _{4,5}^-$
(T$_{1u}$ or T$_{2u}$)
cubic symmetry has five phases near $T_c$.  The cubic group is only
partly broken in each of these phases.  The details are given in 
Ref.\ [3].  The order parameter is $\Delta_k \propto$
$\hat{d}\cdot \vec{k}$, where $\hat{d}$ is the vector labelling each 
phase.
\label{fig-cubPhase}}
\end{figure}


Several experimental signatures for HM AFMs ({\it i.e.} the normal
phase) have been proposed: non-Korringa behavior in NMR \cite{SSS1,SSS2},
resistivity and magnetic susceptibility \cite{Fujii}, 
or more exotically through spin-polarized
angle resolved positron annihilation experiments\cite{hanssen}.
Further tests are possible in the superconducting phase.  
The unconventional pairing of SSS would be reflected in the existence 
of multiple superconducting phases (See Fig.\ 2).
Also, thermodynamic
quantities which have an exponential dependence in ordinary BCS
superconductors will have a power law dependence on the temperature in
SSS due to the nodes of the gap function \cite{SSS2}.  In principle, 
the exponent may be used to distinguish point nodes from line nodes, 
and hence gain symmetry information, but such measurements 
on unconventional superconductors are very demanding of sample quality.
 
Potentially powerful tools to analyze the superconducting
state, including establishing that SSS occurs,
are tunneling and the related phenomena that occur at interfaces.
Single spin superconductivity offers unique tunneling
effects.  Junctions may be formed between two identical SSS
materials, between SSS material and another superconductor (such as
an s-wave superconductor, an unconventional (p-wave or d-wave) 
superconductor or a different SSS) or between a SSS and a normal metal 
(such as an ordinary metal, a ferromagnet, a half-metal or a HM AFM).

Unlike the case of bulk SSS where spin is irrelevant, 
the spin is not only relevant but crucial to SSS junctions.  
Proper treatment of the spin is the new ingredient
required to generalize existing theories of tunneling 
\cite{Likharev,Pals,Fenton,Gulyan} to the single spin
case.  In the bulk of a SSS the order parameter always breaks the
lattice symmetry and often can be characterized by a vector $\hat d$.
In addition, magnetocrystalline anisotropy (although involving 
energies of only $\sim$ 10$^{-4}$ eV) determines the direction of the
atomic moments, and hence the preferred direction of the spin of the
superconducting carriers.  Spin-orbit coupling also induces a non-zero
orbital moment, but we neglect that complication in this paper.
Its effect should be much less important in the double perovskite
HM AFMs than in the heavy fermion unconventional superconductors.

For definiteness we consider a model of SSS junctions based on
the tunneling hamiltonian
\be
H_T = \sum _{\vec{k},s;\vec{p},s\p} 
\left(
T_{\vec{k} \vec{p}} \delta _{s s\p} 
C_{\vec{k}s}^{\dag} C_{\vec{p} s\p} + {\mathrm{h.c.}}
\right)
\label{tunHam}
\ee
where $\vec{k}$ and $s$ are the wave vector and spin indices on the left
side of the junction and $\vec{p}$ and $s\p$ are the corresponding
quantities on the right side.  We assume the basic spin-diagonal form 
$\delta _{s s\p}$ of the tunneling matrix element.
The same coordinate systems are used on both
sides of the junction in Eqn.\ (\ref{tunHam}), however, the orientation
of the spin axes may differ on the two sides of the interface.
Neighboring crystals, or neighboring domains in a monocrystal, may
have non-aligned carrier spins.  
It is convenient to transform to a spin system that 
diagonalizes the local hamiltonian (so that the gapped channel is
purely spin down).  The tunneling hamiltonian becomes
\be
H_T = \sum _{\vec{k},s;\vec{p},t} 
\left[
T _{\vec{k} \vec{p}}  \,
\left( e^{i\vec{\theta} \cdot \vec{\sigma}} \right)_{s t} \,
C_{\vec{k}s}^{\dag} C_{\vec{p}t} + {\mathrm{h.c.}}
\right]
\label{Ht}
\ee
where $t$ is the natural spin coordinate on the right.
The spins $s$ and $t$ differ by a rotation by the angle $\theta$
about the axis $\hat{\theta}$.
$\vec{\sigma}$ denotes the Pauli spin matrices.  

The down spin electrons do not participate in tunneling since the
gap is large compared to the temperature and the junction voltage, 
$E_g \gg kT, eV$.  The spin up part of the tunneling
hamiltonian is simply
\be
H_T = \sum _{\vec{k},\vec{p}} 
\left(
T _{\vec{k} \vec{p}} \,
d^{\half}_{\half , \half} (\theta) \, 
C_{\vec{k}}^{\dag} C_{\vec{p}} + {\mathrm{h.c.}}
\right)
\label{Htup}
\ee
where
$d^j_{m_1,m_2} (\theta )$ is the Wigner d-function \cite{Wigner}.
Apart from the restriction to spin up
electrons, this form of the tunneling hamiltonian is very similar
to the familiar one from BCS theory.  The single-particle tunneling
current may be computed using standard techniques \cite{Likharev}
\begin{eqnarray}
I_S & = & e \, d^{1}_{1,1} (\theta) 
\sum _{\vec{k},\vec{p}} | T _{\vec{k} \vec{p}} | ^2  \\ \nonumber
& & \times \int \! \frac{d\epsilon}{2\pi}
A_R(\vec{k},\epsilon ) A_L(\vec{p},\epsilon + eV )
\left[ n_{F\uparrow} ( \epsilon ) - n_{F\uparrow} ( \epsilon + eV ) \right]
\end{eqnarray}
where $A_L$ and $A_R$ are the spectral functions for the two sides of
the interface.  The spin-1 Wigner d-function is given by 
$d^{1}_{1,1} (\theta) = \half ( 1 + \cos (\theta ) )$.  Similarly,
the Josephson current is 
$I_J = 2e \, \Im [ e^{-2ieVt/\hbar } \Phi _{ret} (eV) ]$
where the Green's function is given by (using standard notation)
\begin{eqnarray}
\Phi _{ret} (i\omega ) & = & d^{1}_{1,1} (\theta) \, 
\sum _{\vec{k},\vec{p}} 
\frac{\Delta _L \Delta _R}{4}
\frac{ T _{\vec{k} \vec{p}}  T _{-\vec{k}, -\vec{p}} }{E_kE_p}
\left\{ \rule[-3mm]{0mm}{7mm}
[1-n_{F\uparrow} (E_p) 
\right.  \nonumber \\
& & 
-n_{F\uparrow} (E_k)] 
\left( 
\frac{1}{i\omega + E_p + E_k} -
\frac{1}{i\omega - E_p - E_k} 
\right)  \nonumber \\
& &  +[n_{F\uparrow} (E_k) -n_{F\uparrow} (E_p)] 
\left( 
\frac{1}{i\omega + E_k - E_p} \right. \nonumber \\
& & 
\hspace{3cm} \left. \left. - \frac{1}{i\omega + E_p - E_k} \right)
\right\}
\end{eqnarray}
where $\Delta _{L,R}$ is the gap function.

The expressions given above may be used to describe many different
SSS junction effects.  Due to space limitation we will focus on
a few of particular interest.
Consider the case of a SSS-SSS$\p$ junction.  
Tunneling is allowed, but only to the extent that a carrier can propagate 
once its axis of spin quantization projects onto the axis of 
the neighboring HM material.
For example, the DC critical current for tunneling between two SSS regions
with $\hat{d}$=(1,i,0) of $\Gamma _4^-$ (T$_{1u}$) cubic gap symmetry is 
given by
\be
I_J^{V=0}  = \frac{\sigma _0}{e} \frac{\pi \Delta}{4}  
\left( \cos (\theta ) + 1\right) \, \sin (\phi ) \, \cos (\varphi ) \,
\, \tanh ( \frac{\beta}{2} \Delta )
\ee
where $\Delta$ is the RMS gap and $\phi$ is the phase
difference of the order parameter.  The maximal normal conductance is
$\sigma _0 = 2 \pi e^2 N_L N_R | T_0 |^2$ where 
$N_{L,R}$ is the density of states and 
$T_0$ is the magnitude of the tunneling matrix element.
Note that the angle $\varphi$, the relative orientation of the order 
parameter on the two sides of the junction, can have a large effect
on Josephson tunneling.  Even if the superconducting state has the same 
symmetry on both sides, the tunneling goes to zero when $\varphi = \pi/2$.
The critical AC Josephson current has the same angle dependence,
and it oscillates at the classic frequency $\nu _J = 2 e V /h$.  
Also, the superconducting
density of states is given by the usual expression
\be
\varrho (eV) = \left( \frac{dI}{dV} \right) _{\! SN} / 
\left( \frac{dI}{dV} \right) _{\! NN} 
\ee
which is independent of the orientation.

In other cases the spin projection factor may cause the second order Josephson 
coupling to vanish entirely.  The archetypical example is the coupling between
an s-wave (BCS) superconductor and a half-metal, normal or superconducting.
The states of different total spin are orthogonal.
Physically, it is impossible for the down spin to tunnel into the
half-metal, unless enough energy is supplied to overcome either 
the gap in the s-wave superconductor or the gap in the down spin
channel of the half-metal.  As a result, the half metal (or SSS) should
behave as if it were an insulator from the point of view of the 
s-wave superconductor.  
This effect has been discussed in the context of triplet-singlet
junctions \cite{Pals,Fenton,Gulyan,Wolf}.

Similarly, there should be no Josephson tunneling
to any unconventional superconductor with pure spin singlet pairing.
Since tunneling is allowed from a single spin to p-wave superconductor
in either the Anderson-Brinkman-Morel or the Balian-Werthamer state,
the SSS Josephson effect could, in principle, be used as a litmus test to 
distinguish p-wave from d-wave pairing.  That is, the knowledge that
half-metallic materials must form single spin pairs (and not d-wave
pairs) makes them a kind of standard by which to probe other unconventional
superconductors.


Finally, we note that the Josephson coupling between SSS and p-wave
superconductors suggests the possibility of induced superconductivity.
In particular, it has been suggested recently that Sr$_2$RuO$_4$ may be
displaying triplet superconductivity \cite{SrRuO}.  
If true, its lattice is similar
enough to those of the candidate HM AFMs that a good interface may 
be possible.  Then it may be possible to use Sr$_2$RuO$_4$ to 
induce SSS.


\vspace{5mm}

\noindent
{\bf Acknowledgments:}

Supported by the Office of Naval Research.

\vspace{3mm}

\noindent
$^{\dag}$E-mail: ~~rudd@dave.nrl.navy.mil

\noindent
$^{\ddag}$E-mail: ~~pickett@physics.ucdavis.edu

\end{document}